\documentclass[12pt, twocolumn]{article}

\usepackage{url}
\usepackage{graphicx}
\usepackage{float}
\usepackage{booktabs}
\usepackage{amsmath}

\begin{document}

\title{A Bayesian binary algorithm for RMS-based acoustic signal segmentation}
\author{Paulo Hubert$^1$ \and Alexandra Chung$^1$ and Linilson R. Padovese$^1$}

\date{$^1$ Department of Mechanical Engineering, Escola Politecnica - University of S\~{a}o Paulo, S\~{a}o Paulo, SP - Brazil \\
\today} 


\maketitle

\abstract{
Changepoint analysis (also known as segmentation analysis) aims at analyzing an ordered, one-dimensional vector, in order to find locations where some characteristic of the data changes. Many models and algorithms have been studied under this theme, including models for changes in mean and / or variance, changes in linear regression parameters, etc. In this work, we are interested in an algorithm for the segmentation of long duration acoustic signals; the segmentation is based on the change of the RMS power of the signal. We investigate a Bayesian model with two possible parameterizations, and propose a binary algorithm in two versions, using non-informative or informative priors. We apply our algorithm to the segmentation of annotated acoustic signals from the Alcatrazes marine preservation park in Brazil.


}

\maketitle

\section{Introduction} \label{sec:intro}

The problem of signal segmentation arises in different contexts \cite{Makowsky2014, Ukil2006, Schwartzman2011, Kuntamalla2014, Theodorou2014}. The problem is broadly defined as follows: given a discretely sampled signal $y \in \Re^N$, divide it in contiguous sections that are internally homogeneous with respect to some characteristic. The segmentation is thus based on the premise that the signal structure changes one or many times during the entire sampled period, and one is looking for the times where the changes occur, i.e., the \emph{changepoints}. 

In this work we are interested in segmenting acoustic signals, more specifically underwater acoustic signals acquired off the Brazilian coast. Since 2010, the Acoustics and Environment Laboratory (LACMAM) at University of S\~{a}o Paulo has been designing equipment for underwater acoustic monitoring \cite{CaldasMorgan2015}; and from the past few years, we have acquired and stored over 2 years of acoustic recordings taken from different locations, amounting to more than 15 Tb of data.

The main challenge in exploring these data lies on the abundance of interesting events, and at the same time on the sparsity of such events. The sparsity of events makes the direct inspection of long duration signals a very demanding task, while the variety of potentially interesting events discourages the design and application of detection algorithms aimed at specific events, for they would potentially miss many unexpected (and for this exact reason, interesting) events.

We are currently developing an unsupervised learning approach, based on the tripod \emph{segmentation, characterization and categorization}, to deal with this situation. The idea is to first divide the long duration signal into sections which are likely to contain different sets of events; then, we characterize each section by using a sparse representation approach, and finally we cluster the segments together or categorize then in a sequential manner. 

This paper deals with the first task of the tripod: the segmentation of the signal. Our approach is based on the hypothesis that the occurrence of an event induces an immediate change on the total sound pressure level, and that this change can be detected on the variance of the signal's amplitude. What we seek then is a variance changepoint detection algorithm. 

A few algorithms to detect changes in signal's variance are available; in the next section we give a quick review on both the signal segmentation and changepoint analysis literatures. After that, section \ref{sec:seqseg} defines the algorithm to be used for the segmentation; section \ref{sec:res} presents our results in the segmentation of both simulated and real acoustic signals, and section \ref{sec:conc} concludes the paper.

\subsection{Changepoint analysis and signal segmentation }\label{sec:cpsigseg}

Even though the problems of changepoint analysis and signal segmentation are very closely related, the literatures adopting each nomenclature are somewhat independent.

As for the signal segmentation literature, both probabilistic and non-probabilistic methods can be found, see \cite{Theodorou2014} for an interesting review. These algorithms have a few features in common:

\begin{enumerate}
 \item The use of a more or less detailed parametric model to describe the signal;
 \item The definition of frames, or windows, to characterize local behavior;
 \item A peak detection or thresholding procedure applied to the collection of frames to obtain segments' boundaries.
\end{enumerate}

These methods are well suited for the analysis of short to medium term signals (up to a few thousand data points), because the estimation step for the parametric models, be it a discrete Fourier or wavelet transform, and / or a filtering procedure, is usually computationally intensive. Also, the use of a detailed parametric model is adequate only when the additional structure imposed by the model over the original signal is well justified, i.e., when the phenomena causing the change in the signal's characteristics is reasonably well known.

The changepoint literature, in the other hand, is more prolific and has more of a statistical flavor to it; see \cite{Shaban1980} for a review on changepoint research up to the decade of 1970.

In the changepoint literature, the problem is modelled over a one-dimensional signal (a real or complex vector) obtained from the noisy measurements of some system. The properties of the system change over time, altering the signal in a measurable way. There are two main cases of this problem: 1) the goal is to detect a change and act immediately; this is usually called \emph{realtime} segmentation; and 2) the goal is to analyze a long pre-recorded signal and find all the changepoints in it, along with estimatives of the system's state inside each block; this is usually called \emph{retrospective} segmentation.

The recent literature proposes a few solutions for the problem. \cite{Jackson2005}, for instance, provides a general method based on dynamic programming that is able to find the global optimum of a fitness function, $V(P) = \sum g(B_m)$, where the sum is taken over $m$ blocks, and $g$ is the fitness function of a single block (usually a likelihood based on a probabilistic model), in  $O(N^2)$  time. 


In the same spirit, \cite{Killick2012} improves the work of Jackson by proposing a \emph{Pruned Exact Linear Time} (PELT) algorithm that, under mild conditions, is able to optimize the global fitness function with complexity $\mathcal{O}(n)$. Killick's method is general, and can be applied to any fitness function that fulfills a mild condition on the relation between the fitness of an entire segment and the fitness of the same segment divided by one changepoint (for details, see the original paper \cite{Killick2012}).

Many other papers are available on the subject, both under the names of segmentation and changepoint analysis. We intend to write a second paper offering a compared review of the two approaches for the problem, but for now our main goal is to present a new, Bayesian binary algorithm, that is closer in spirit to the methods found in the changepoint literature. Our algorithm approaches the problem of segmentation as one of sequential hypothesis testing. We adopt a binary strategy, first finding the best changepoint for the entire signal, and, if this changepoint is accepted, applying the procedure recursively to each segment obtained. In the next section, we define our model and the Bayesian binary algorithm.

\section{A Bayesian algorithm for variance changepoint detection} \label{sec:seqseg}

We start by assuming that the (discretely sampled) signal at time $t$, $y_t \in \Re$, has $0$ mean amplitude for all $t$, and finite power $\sigma_t^2$. We adopt a Gaussian probabilistic model for the signal, $y_t \sim \mathcal{N}(0, \sigma_t^2)$. The choice of the Gaussian model can be justified by the maximum entropy principle \cite{Jaynes1982, Jaynes1987}, which states that the most conservative probabilistic model to be adopted in any situation is the one which maximizes Shannon's entropy $H(p) = -E\left(log \: p\right)$ (where $p$ is the model's density, and the expectation is taken with respect to $p$) conditionally on what we already know about the data (in this case, $0$ mean amplitude and finite variance). This maximization of entropy guarantees that we are not allowing any hidden assumptions into our model, and this kind of reasoning can keep the algorithm more robust to deviations from the model's assumptions, as we will see later on.

We will assume that $\sigma_t^2$ is a piecewise constant function on $t$, and we are interested in estimating the localization of discontinuities or jumps in this function.

\subsection{Binary algorithms}

One of the simplest ways to tackle the changepoint location task is by using a binary algorithm. Given the entire signal, the first part of the algorithm looks for the single changepoint that is most likely or best in some sense. After obtaining this changepoint, the traditional binary approach will apply the same procedure recursively to the newly obtained segments. The stopping condition is usually based on a model selection criteria.

Our algorithm differs from the traditional binary strategy in that it will apply a statistical hypothesis testing procedure at each step to decide if a given changepoint is valid (i.e., if there is enough evidence in the data that there is indeed a change at this point). If the changepoint is considered valid, the algorithm continues to  estimate new changepoints in the two segments obtained from the last iteration. If not, the execution is halted.

The binary segmentation algorithm is then based on a single changepoint model defined as follows:

\begin{equation}\label{eq:model}
y_t \sim
  \begin{cases}
   \mathcal{N}(0, \sigma_0^2)&\quad\text{if } t \le \bar{t}\\
   \mathcal{N}(0, \sigma_1^2)&\quad\text{if } t > \bar{t}\\
  \end{cases}
\end{equation}

The likelihood function associated with this model is thus
\begin{equation}\label{eq:verot}
\begin{aligned}
 \mathcal{L}(\bar{t}, \sigma_0^2, \sigma_1^2 | y)& = (2\pi\sigma_0^2)^{-\frac{\bar{t}}{2}}(2\pi\sigma_1^2)^{-\frac{N-\bar{t}}{2}} \times \\
 & exp\left[-\frac{\sum_{t=1}^{\bar{t}}y_t^2}{2\sigma_0^2}-\frac{\sum_{t=\bar{t}+1}^Ny_t^2}{2\sigma_1^2}\right]
\end{aligned}
\end{equation}

The first part of the algorithm involves picking the best value for $\bar{t}$; in so doing, the values of $\sigma_0$ and $\sigma_1$ are not important, i.e., they are \emph{nuisance} parameters. To eliminate this parameters and obtain the marginal posterior of $\bar{t}$, we choose priors for each parameter and integrate them out.

For variance parameters like $\sigma_0$ and $\sigma_1$, it is well-known in the Bayesian inference literature that to obtain an uninformative prior one should not adopt the usual uniform distribution for $\sigma$, but rather an uniform for $log(\sigma)$, the so-called Jeffreys' prior \cite{Jaynes1968, Jeffreys1946} $\pi(\sigma_0) = 1/\sigma_0$ for $\sigma_0$. These priors, besides being uninformative and invariant to different parameterizations of the model (over variances or precisions, for instance), allow analytical integration of equation \ref{eq:model}, yielding the marginal posterior
\begin{equation}\label{eq:postt}
 \begin{aligned}
 P(\bar{t} | y) \propto & \pi(\bar{t})\cdot\left(\sum_{t=1}^{\bar{t}}y_t^2\right)^{-\frac{\bar{t}}{2}} \left(\sum_{t=\bar{t}+1}^Ny_t^2\right)^{-\frac{(N-\bar{t})}{2}} \times \\
 & \Gamma\left(\frac{\bar{t}}{2}\right)\Gamma\left(\frac{N-\bar{t}}{2}\right)
 \end{aligned}
\end{equation}

With this posterior, the algorithm now must estimate the best unique changepoint for the current segment. This is a standard statistical estimation procedure, and as is well-known, different cost functions to evaluate the estimation error yield different estimators. If the cost function is quadratic, the best changepoint is the posterior mean; if the cost function is the absolute value, the best changepoint is the median, if the cost function is a $0-1$ function, the best changepoint is the posterior mode.

In this algorithm, neither the median nor the mean estimator would be ideal, specially because the assumption of a single changepoint is most likely false. Consider, for example, figure \ref{fig:post_t} below, that shows the single changepoint posterior calculated on a signal with two changepoints.

\begin{figure}[H]
 \centering
 \includegraphics[width=0.4\textwidth]{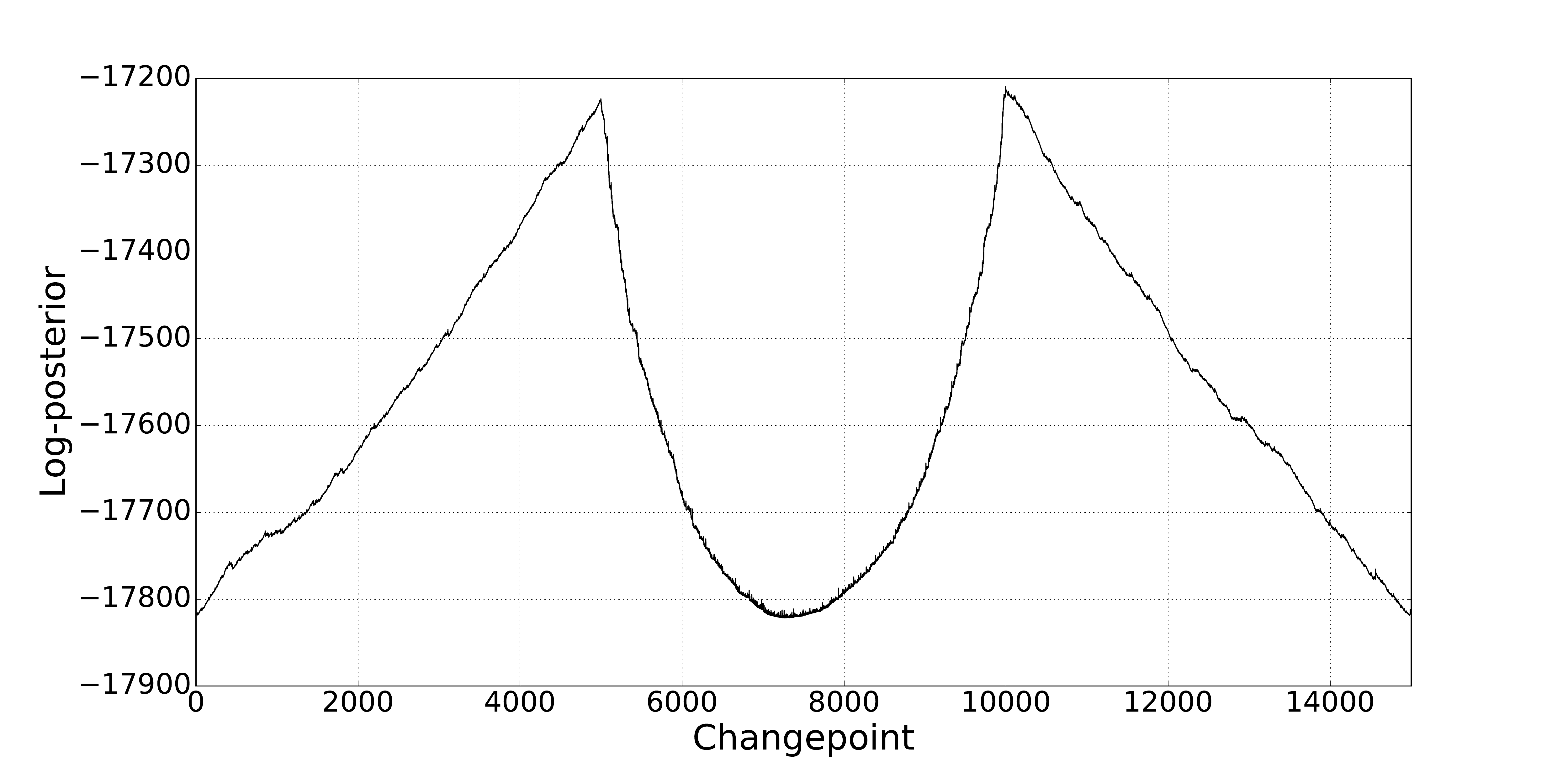}
 \caption{Posterior for a single changepoint using data with two changepoints}
 \label{fig:post_t}
\end{figure}

Both the mean and median of this distribution are located near the center, which is not close to neither changepoint. The posterior mode, however, is robust to the number of changepoints being greater than $1$, and this will be the estimator of choice.

This choice defines the first part of the algorithm: obtain the marginal posterior, and its mode. The discrete optimization involved in the determination of the posterior mode can be carried out by direct inspection, which can be parallelized.

In the next step of the algorithm, the goal is to determine the validity of the changepoint based on the evidence that the data gives about this changepoint being a true one. 

\subsection{Full Bayesian evidence measure}

To be a valid changepoint, in the present context, means that the signal variances of the two segments are different. So this step requires an equality of variances test.

From the full model's likelihood \ref{eq:verot}, conditioning on $\bar{t}$ and multiplying by the joint prior on $(\sigma_0, \sigma_1)$ yields the posterior

\begin{equation}\label{eq:postfull}
 P(\sigma_0, \sigma_1 | y, \bar{t}) \propto \pi(\sigma_0, \sigma_1)\cdot \mathcal{L}(\bar{t}, \sigma_0^2, \sigma_1^2 | y)
\end{equation}

This time, however, it is obviously not desirable to marginalize out $\sigma_0$ and $\sigma_1$, since now these parameters are no longer nuisant. They are, in fact, the very parameters that must be tested for equality: $H_0: \sigma_0 = \sigma_1$ is the hypothesis of interest.

It is important to note that the full model \ref{eq:postfull} is defined over a $2$-dimensional parametric space, and that $H_0$ describes a lower ($1$-)dimensional manifold on this original space. Hypothesis that define lower dimensional manifolds on the parametric space are called \emph{sharp} or \emph{precise} hypothesis in the Bayesian literature \cite{Dickey1970}. 

These hypothesis are challenging to test in the usual Bayesian hypothesis testing frameworks, because the posterior measure over $H_0$ is by definition $0$. However, in \cite{Stern1999}, an evidence measure for sharp hypothesis is presented; this measure is shown to be fully Bayesian (in the sense that it arrives directly from a particular cost function \cite{Madruga2001}), and to possess many desirable properties. The literature presents already many situations where this measure was succesfully applied \cite{Hubert2009, Diniz2012, Chakrabarty2017, Hubert2017} to sharp hypothesis settings in different problems.

Following the original authors, we call this measure the \emph{e-value}, $ev(H_0)$ being the evidence value in favor of $H_0$. 

The full definition and analysis of the e-value is beyond the scope of this paper; the interested reader is directed to the previously cited references, in special \cite{Stern1999}. However, to keep this work reasonably self-contained, we now define the e-value in broad terms.

Given a full posterior model $P(\theta | x)$ with $\theta \in \Theta$, and given a sharp hypothesis $H_0: \theta \in \Theta_0$ with $dim(\Theta_0) < dim(\Theta)$, obtain the maximum value of the full-posterior restricted to $\Theta_0$

\begin{equation*}
\begin{aligned}
 & \theta^* = arg max_{\theta \in \Theta_0} P(\theta | x) \\
 & p^* = P(\theta^* | x)
\end{aligned}
\end{equation*}

Now define the tangent space or surprise set as

\begin{equation}\label{eq:surprise}
 T(p^*) = \left\{ \theta \in \Theta : P(\theta | x) > p^* \right\}
\end{equation}

The tangent space is the set of all parameter values with higher posterior density than the maximum posterior under $H_0$. If this set has high posterior measure, it means that $H_0$ does not traverse regions of high posterior density, and the evidence in favor of $H_0$ must be low. In fact, define

\begin{equation}\label{eq:ev}
 ev(H_0) = 1-\int_{T(p^*)} P(\theta | x) d\theta
\end{equation}

to be the evidence in favor of $H_0$. The evidence will take the value $0$ if the measure of the surprise set is $1$ (i.e., if the maximum posterior value under $H_0$ is almost surely the minimum unrestricted posterior value), and conversingly the evidence in favor of $H_0$ will be $1$ if the measure of the surprise set is $0$ (i.e., the maximum posterior under $H_0$ is almost surely the unrestricted maximum). 

As the definition above shows, the calculation of the e-value involves two steps: an optimization step and an integration step. The optimization is constrained to $\theta \in \Theta_0$, and will depend on the choice of priors; sufficiently simple priors will lead to analytical solutions to this step.

The integration step can be carried out by Markov Chain Monte Carlo methods, as is usual in Bayesian inference procedures.

\begin{figure}[H]
 \centering
 \includegraphics[width=0.4\textwidth]{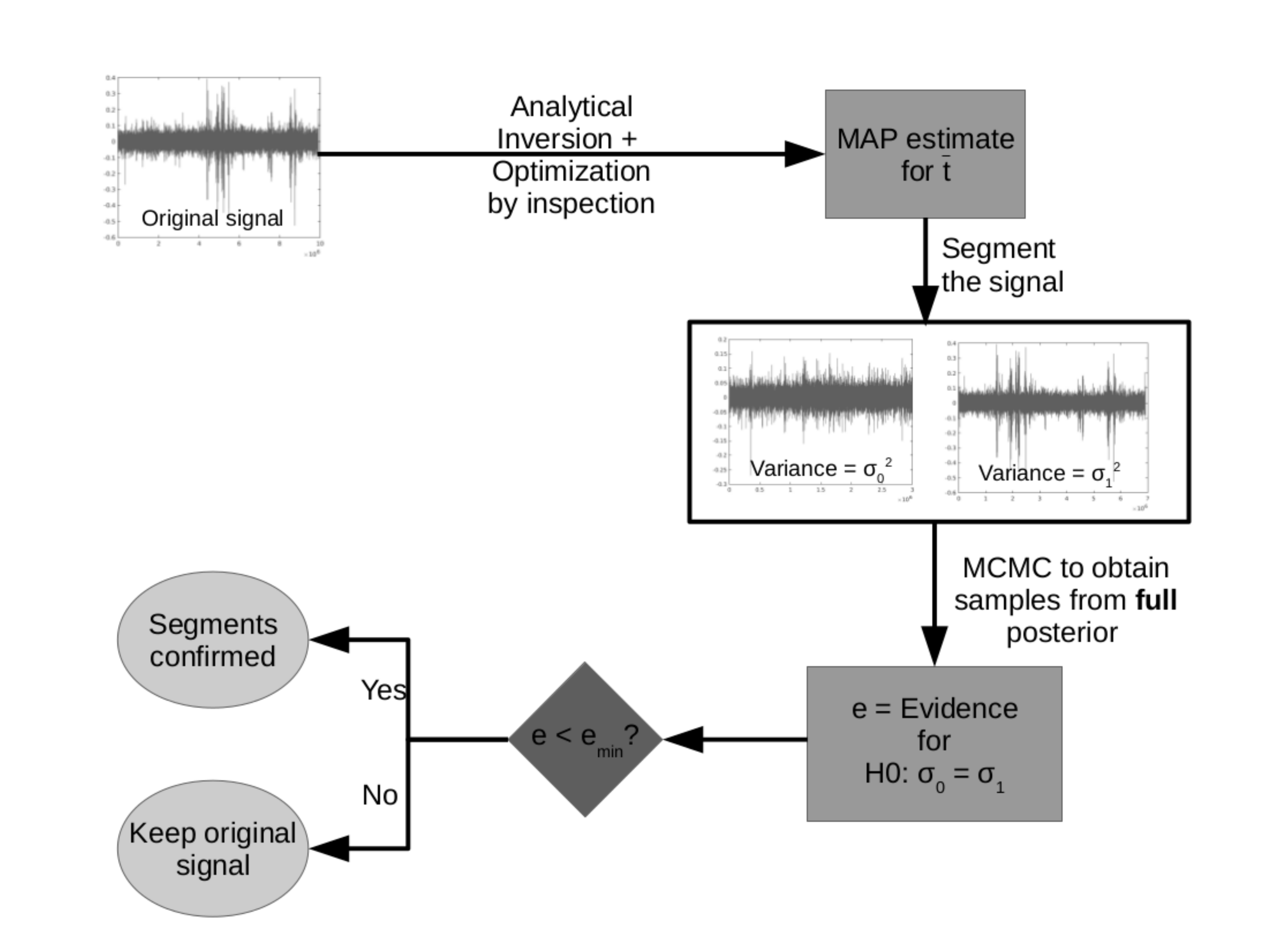}
 \caption{One step of the sequential segmentation algorithm.}
 \label{fig:diag}
\end{figure}

This finishes the definition of the binary algorithm. One full step of the algorithm will consist of two substeps: first, to estimate the segmentation point $\bar{t}$; second, to compare the variance of the segments, calculating a measure of evidence for the hypothesis $H_0:\sigma_0 = \sigma_1$. A diagram illustrating the algorithm's flow can be seen in Figure \ref{fig:diag}.

\subsection{Priors and the power of the e-value}

To calculate the e-value in the segmentation model \ref{eq:postfull}, all that is left to do is to pick a joint prior $\pi(\sigma_0,\sigma_1)$, and from then on follow the procedure delineated above.

One obvious choice for the priors is to adopt the product of Jeffreys' priors $(s_1s_2)^{-1}$; by doing so, the model is treating both these parameters as completely unknown in advance, i.e., the algorithm will act as if it knows nothing about the segments' variances and the relation between them.

This choice gives the optimal value 

\begin{equation}\label{eq:optsig}
 \sigma_* = \frac{\sum_{t=1}^Ny_t^2}{N + 2}
\end{equation}

for the signal's variance under $H_0$ (no changepoint). To calculate the evidence in favor of $H_0$, we estimate the integral of the posterior over the surprise set by the adaptive MCMC method of \cite{Haario2001}.

\begin{figure}[H]
 \centering
 \includegraphics[width=0.4\textwidth]{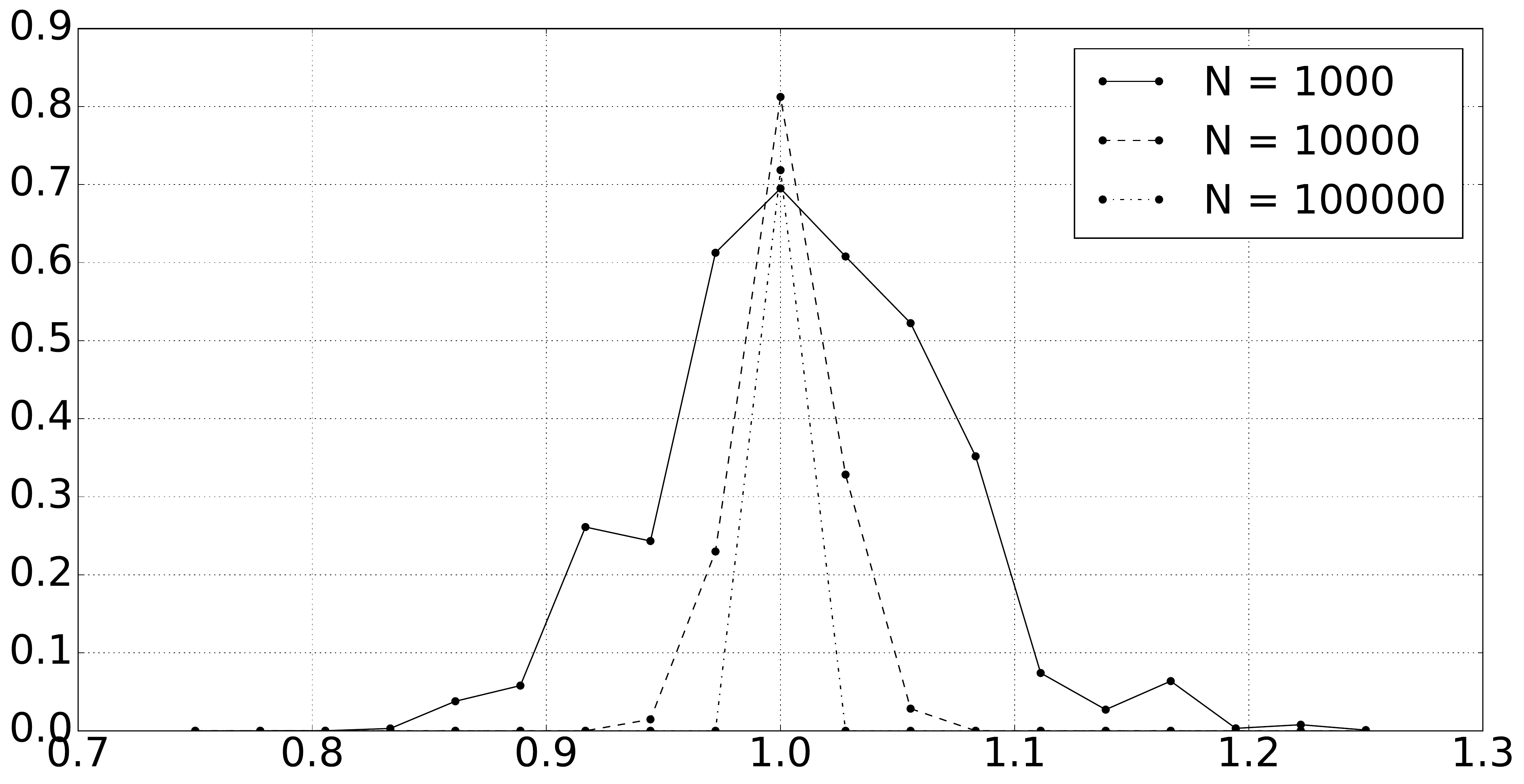}
 \caption{Evidence value for $H_0$}
 \label{fig:evsim}
\end{figure}

To verify the behaviour of the e-value with this choice of priors, we simulate Gaussian signals with various sample sizes, divided into two segments, with the variance of the first segment set to $1$, and that of the second segment varying in $[0.5, 1.5]$. Figure \ref{fig:evsim} shows the evidence in favor of $H_0$ for several values of $\sigma_1$ and several sample sizes.

\begin{figure}[H]
 \centering
 \includegraphics[width=0.4\textwidth]{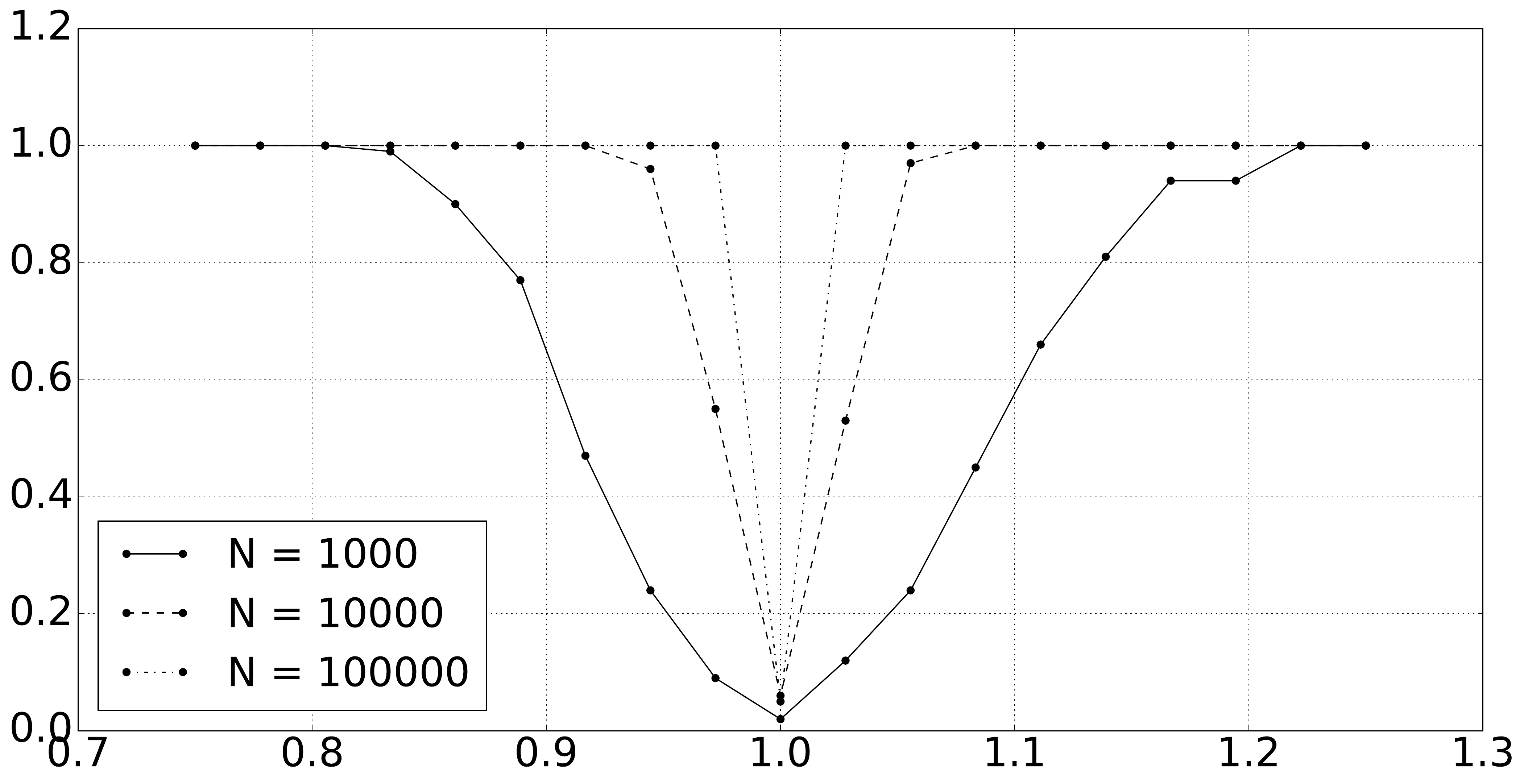}
 \caption{Power of the test based on the e-value}
 \label{fig:evpower}
\end{figure}

It is very important to take notice that the e-value is \emph{not} a significance measure, i.e., it does not result from a \emph{control type-I error} procedure. This implies that the sampling distribution of  the e-value is not uniform; however, a transformation exists that changes the e-value into a significance measure \cite{Stern2008}. Using this transformation, it is possible to fix the type-I error at $0.05$ and evaluate the power of the test. The result for different sample sizes and values of $\sigma_1$ is on figure \ref{fig:evpower}

\subsection{Using informative priors}

The test based on the (transformed) e-value is quite powerful, as the simulations indicate. The power, as expected, gets higher for greater sample sizes; this means that the test will detect smaller deviations from $\sigma_0 = \sigma_1$ as the sample size grows, while at the same time keeping the type-I error probability fixed. 

This is an important issue, specially in the segmentation algorithm where the test will be sequentially applied to the comparison of segments with different sample sizes. If we choose to keep $\alpha$ (probability of type-I error) fixed, the power of the test will change as the sample size changes. However, in a signal detection setup, usually one desires to balance both type-I and type-II error probabilities regardless of the size of the incoming signal.

The relation between significance levels, test power and sample size is a deep and often discussed question in hypothesis testing \cite{Pericchi2016, PereiraPrint}. Recent literature proposes to change the significance level as the sample size changes, to keep some relation $u(\alpha,\beta)$ between the probabilities of both error types at a constant value. This can be done by using adaptive significance levels (given by a function of the sample size $n$, see \cite{Pericchi2016}) or by imposing an ordering on the parameter space based on Bayes factors \cite{PereiraPrint}. 

Usually, the procedure starts by asking the researcher to pick a sensibility, and the type-I error probability for the test given a value for $n$. After that, the statistician calculates the respective power of the test, and obtains a rule to define the new significance value for a new value of $n$, in order to keep constant the relation $u(\alpha, \beta)$. 

For the segmentation task, however, and in our particular application (segmentation of large samples), the algorithm will have to work with segments of very different sizes (from $~10000$ to more than $9$ million), and the adaptive significance level would also vary wildly. The consequence is that, for the larger segments, the algorithm would require very small significance values; and in a MCMC setting, higher precision for the probability estimates means longer chains, and longer chains mean higher execution times. 

So instead of using an adaptive significance value, we propose instead to use a strongly informative prior, and use the hyperparameters to calibrate the power of the procedure.

This idea was first introduced in a previous paper \cite{Hubert2018}. The paper analyzes the binary algorithm for signal segmentation, but uses a different parameterization $\theta = (\sigma_0, \delta)$ where $\delta = \sigma_1 /\sigma_0$. Independent priors for these two parameters are proposed, one that is uninformative on the value of $\sigma_0$, and strongly informative over $\delta$. The advantage of working with $(\sigma_0, \delta)$ instead of $(\sigma_0, \sigma_1)$ is that $\delta$ is a pure number, i.e., it does not depend on scale. It can be interpreted as the quotient between the power of any two contiguous segments.

There are however some difficulties in working with $\delta = \sigma_1 / \sigma_0$, one of them being that $\delta$ must be nonnegative. For this new, current version of the algorithm, we parameterize the problem using $\delta = log(\sigma_1 / \sigma_0)$, and propose a Laplace prior with the form

\begin{equation}\label{eq:laplace}
 p(\delta) = \frac{1}{2\beta}e^{-\frac{|x|}{\beta}}
\end{equation}

The above Laplace distribution has a peak on $x = 0$, and the peak is sharper as the value of $\beta > 0$ decreases. The Laplace distribution is a maximum entropy prior, i.e., it is the probability distribution with higher entropy subject to the constraint $E(|x|) = \beta$. 

The segmentation algorithm works as above, except that now the e-value calculation uses the Laplace prior for $\delta$. This prior, when $\beta$ is close enough to $0$, changes significantly the power of the test, and thus allows tuning of the algorithm's behavior. 

\begin{figure}[H]
 \centering
 \includegraphics[width=0.4\textwidth]{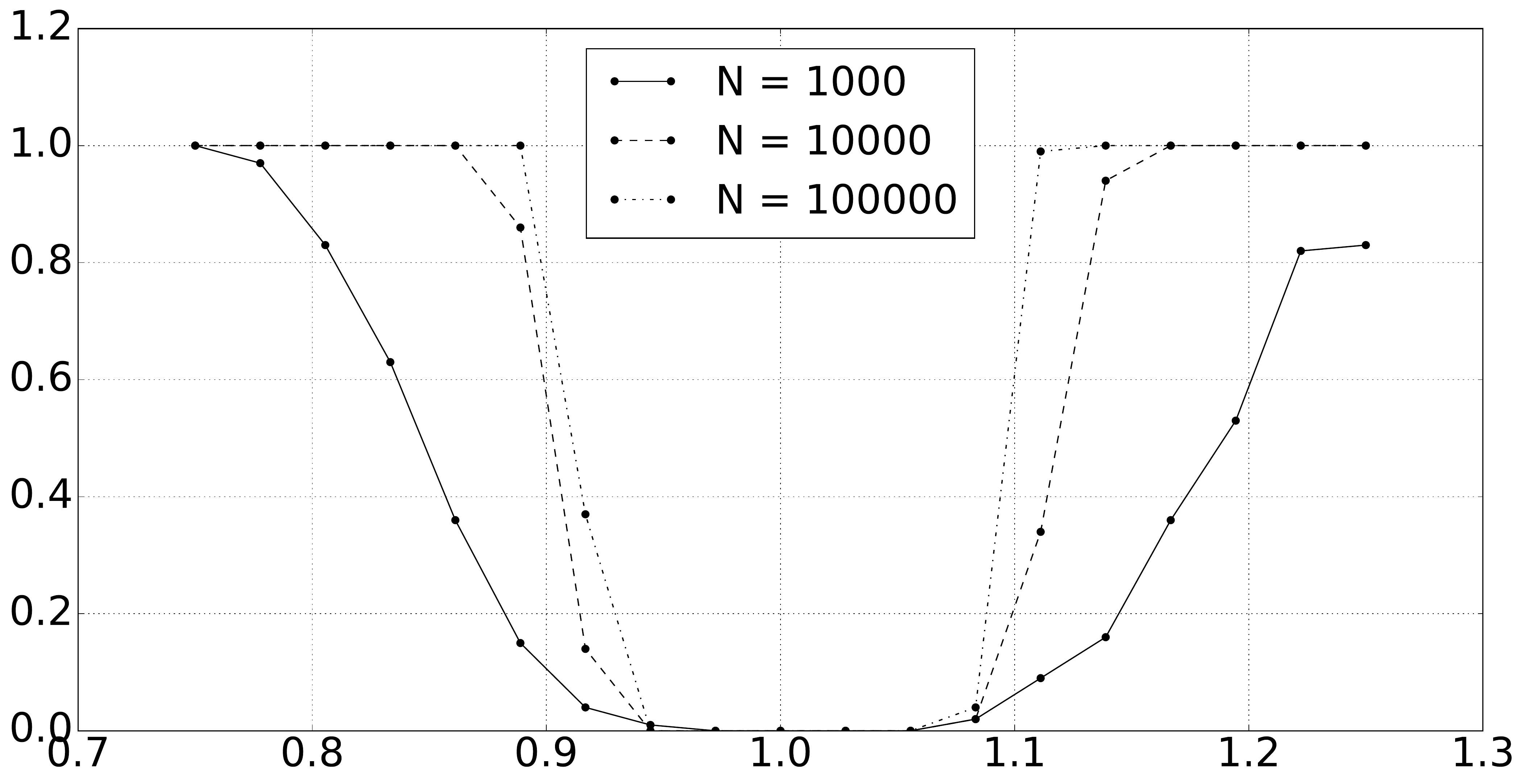}
 \caption{Power of the test based on the e-value with strongly informative priors}
 \label{fig:evpower_beta}
\end{figure}

Figure \ref{fig:evpower_beta} shows the same estimation of power as in figure \ref{fig:evpower}, but this time using the Laplace prior. The values of $\beta$ where taken as $0.005, 0.0005, 0.00005$ for $N=1000, 10000, 10000$ respectively.

Being able to control the power of the test will prove useful when segmenting underwater acoustic signals; in this setting, long segments with true stationary power are not to be expected, even when the segment is capturing a single event. That is the case because both the background noise and the event's physical cause might be changing, due to many factors (including the weather, the movement of event's causes relative to the hydrophone, among others). With a high sampling rate (the data we use in this paper was sampled at $24KHz$) the e-value would give strong evidence against $H_0:\sigma_0 = \sigma_1$ even inside a segment containing a uniform event, and this would lead to oversegmentation. To control the power of the test using an informative prior will allow the algorithm's sensibility to be tuned to the goals of the analysis: if one is interested in capturing larger sections, that might suffer an internal power change that is small compared to the difference between the segment overall power and the background noise power, one only needs to adjust the hyperparameter accordingly.

\subsection{The resolution parameter}

The most demanding step in our binary algorithm is the optimization procedure that looks for the most likely changepoint at each step. This is done by a brute force procedure, that can be parallelized but nevertheless is costly, specially with long signals.

One way to increase the speed of our algorithm is to limit the search for the optimal changepoint: instead of calculating the objective function for all $i \in \{1, ..., N\}$, we can instead calculate the objective only for $i = lj, j \in \{1,...,N/l\}$. 

If the (discrete) posterior for $\bar{t}$, the changepoint parameter, is not very sharp around its maximum, and if the minimum expected segment length is also not too small, $l$ above can be set to a high value, increasing the speed of the algorithm while still being able to identify the most probable changepoints at each step.

However, and since the optimization step will be applied many times, to segments of different lengths, it is not advisable to pick a fixed integer value for $j$; imagine, for instance, that we fix $j = 1000$. In a signal of size $N = 1,000,000$, this value won't stop the algorithm from finding the optimal value (or some good approximation to it); however, for a signal of size $N = 10,000$, it is quite possible that using $j=1000$ will cause the algorithm to miss the optimal point. For this reason, we adopt an adaptive resolution strategy: we pick a starting value for the resolution (say $j = 1000$), but as the algorithm starts obtaining new segments, it will keep the ratio $j/N$ fixed at each step.

%
%
%
%
%

\subsection{The PELT algorithm}

As a basis of comparison to the Bayesian binary algorithm results, we use the PELT algorithm of \cite{Killick2012}; the PELT (\emph{Pruned Exact Linear Time}) algorithm solves the dynamical optimization problem exactly, yielding the global optimum of the model. It does that with $\mathcal{O}(n^2)$ complexity in the worst case, but it can be shown to have $\mathcal{O}(n)$ complexity under mild conditions. 

The algorithm is defined in terms of an additive cost function

\begin{equation}\label{eq:cost}
 C(\{t_i\}) = \sum_{i=1}^{m+1}\left[\mathcal{C}(y_{t_{i-1}+1:t_i}\right] + \beta f(m)
\end{equation}

where in the case of detection of variance changepoints

\begin{equation}\label{eq:pcost}
\begin{aligned}
\mathcal{C}(y_{t_{i-1}+1:t_i}) = & -\frac{|t_j-t_{j-1}|}{2} log \left(\sum_{i=t_{j-1}}^{t_j} y_i^2\right) + \\
 & log\left[\Gamma\left(\frac{|t_j-t_{j-1}|}{2}\right)\right]
\end{aligned}
\end{equation}

and $f(m)$ is the penalty or regularization function for the number of segments. 

The penalty function is essential, since the direct optimization of the cost function will lead to overfitting (which, in this case, will mean oversegmentation). In our tests below, we adopt the MBIC penalty function \cite{Zhang2007}, which is the penalty function used by default by the R package \emph{changepoint} that implements the PELT algorithm \cite{Killick2014}.

For further comparison of our algorithm with other alternatives, we also run the binary segmentation algorithm of \cite{Scott1974}, which is also implemented by the R package \emph{changepoint}.

\section{Results}\label{sec:res}

\subsection{Simulated data}

To analyze the performance of the Bayesian binary algorithm, we start by simulating Gaussian signals with constant mean and variance. We then simulate the changepoint process by using a geometric distribution to model the times between changepoints, and multiply the signal between changepoints for a given factor in order to obtain different variances.

It is clear that the effectiveness of a changepoint detection algorithm depends directly on both the size of the segments, and the magnitude of the jump in the process parameters. To observe the behavior of all algorithms with varying segment sizes, we will keep the expected number of changepoints fixed at $50$ changepoints regardless of the signal's size. When the signal's size $n$ changes, the expected length of the segments will change accordingly (linearly with $n$).

To simulate the magnitude of change in power between segments, we force the segments to alternate variances between $1.0$ and $2.0$. 

The simulation of the changepoint process was repeated ten times for each value of $N$, and we report the average results for each of these values.

The results appear in table \ref{tbl:sim}. The table reports the true number of changepoints in the simulated signal, the estimated total number of changepoints for each algorithm, and the F1 score. The F1 score is calculated as 

\begin{equation*}
F1 = \frac{precision*recall}{precision + recall}
\end{equation*}

where $precision$ is the number of true positives divided by the total number of changepoints identified, and $recall$ is the number of true positives divided by the total number of true changepoints. To accept an estimated changepoint as a true one, it must be between $N/100$ points of a true changepoint.

The value of $\alpha$ for the Jeffreys prior, and the values of both $\alpha$ and $\beta$ for the Laplace prior were selected using the Bayesian Information Criterion (BIC); both the PELT and the BinSeg algorithms utilized the Modified BIC of Zhang \cite{Zhang2007}.

\begin{table*}
\caption{Simulation results; see text for details}
\label{tbl:sim}
\centering
\begin{tabular}{cccccc}

       N & Algorithm &       Time (s) &  True k &  Estimated k &  F1 score \\
\midrule
   10,000 &    binseg &   0.407200 &    34.3 &          2.4 &  0.085693 \\
   10,000 &  jeffreys &   0.210437 &    34.3 &          4.0 &  0.172064 \\
   10,000 &   laplace &   0.245093 &    34.3 &          5.9 &  0.236544 \\
   10,000 &      pelt &   0.037800 &    34.3 &          5.1 &  0.218018 \\
   \midrule
   50,000 &    binseg &   2.151700 &    46.1 &         15.9 &  0.489096 \\
   50,000 &  jeffreys &   1.628161 &    46.1 &         28.6 &  0.701796 \\
   50,000 &   laplace &   1.563547 &    46.1 &         34.1 &  0.761310 \\
   50,000 &      pelt &   0.177500 &    46.1 &         30.7 &  0.793996 \\
   \midrule
  100,000 &    binseg &   4.269800 &    45.9 &         29.5 &  0.772511 \\
  100,000 &  jeffreys &   2.624351 &    45.9 &         37.3 &  0.840989 \\
  100,000 &   laplace &   2.394387 &    45.9 &         41.7 &  0.872812 \\
  100,000 &      pelt &   0.333200 &    45.9 &         38.2 &  0.907438 \\
  \midrule
  500,000 &    binseg &  20.954300 &    50.8 &         42.6 &  0.870825 \\
  500,000 &  jeffreys &   4.558587 &    50.8 &         50.2 &  0.888668 \\
  500,000 &   laplace &   4.088778 &    50.8 &         49.9 &  0.828553 \\
  500,000 &      pelt &   1.997400 &    50.8 &         49.1 &  0.981732 \\
  \midrule
 1,000,000 &    binseg &  20.661000 &    51.8 &         40.0 &  0.372078 \\
 1,000,000 &  jeffreys &   6.243566 &    51.8 &         53.7 &  0.924682 \\
 1,000,000 &   laplace &   5.911876 &    51.8 &         56.5 &  0.921549 \\
 1,000,000 &      pelt &   3.909400 &    51.8 &         50.0 &  0.982603 \\
\bottomrule
\end{tabular}
\end{table*}

The PELT algorithm was the quickest and also the most accurate algorithm on average for all signal sizes, except for $N=10,000$ where the Bayesian binary algorithm with the Laplace prior showed a higher F1 score. The binary algorithm of Scott \cite{Scott1974} was always the slowest and less precise; also, since it is implemented recursively, for longer signals there was an operational system error related to the stack size that stopped the algorithm from running in many simulations.

The Bayesian binary segmentation can be seen to be competitive with PELT in both execution time and accuracy. The use of an informative (Laplace) prior improved the accuracy in almost all scenarios.

In the next section, we apply the Bayesian binary algorithm and PELT to real underwater acoustic signals; the binary algorithm won't be tested because it is unpractical for signals of the size we will be using.

\subsection{Underwater acoustic signals}

Now we apply the four algorithms to the segmentation of real underwater acoustic signals. These signals were obtained by the LACMAM's team on 2017, in the region of Alcatrazes, an archipelago $35$ km off the Brazilian coast, in the city of S\~{a}o Sebasti\~{a}o, SP. More information about the data and the experiment can be found in \cite{SanchezGendriz2017}.

One of the main goals in acquiring these samples is the study of acoustical signatures of boats. Alcatrazes is a marine ecological reserve, the second largest in Brazil, and as such fishing is prohibited in the archipelago's area. As passive acoustic monitoring is cheap, efficient algorithms for boat detection using hydrophone data are a valuable resource to the reserve's fiscalization authorities. 

The laboratory has, by January, 2019, collected almost two years of acoustic signals from the reserve's region. In these signals, many events can be found: the passage of boats, but also fish and whales' vocalizations, and other events with both biological and anthropogenic sources. These events, however, are scarce, making the direct inspection and annotation of the signal a demanding task. 

The segmentation algorithm will be used to aid in this inspection, by first separating sections of the signal that are likely to contain any significant event. 

To test the segmentation algorithms, we have chosen two $15$ minutes long samples where visual inspection of the spectrogram shows many short duration events. After examination of the spectrograms, the samples were listened to and the start and finish times of all events were annotated. A total number of $32$ changepoints were detected, all of them caused by the passage of boats. What we expect is that the segmentation algorithm will be able to correctly identify the boundaries of these events.

One disclaimer is due at this point. The inspection of the samples was aimed at the separation of samples of the acoustic signal generated by the passage of boats. The researcher responsible for the annotation, thus, was not looking to annotate changes in the signal power. For that reason, it is not expected that any algorithm will get high measures of precision or recall. 

The sampling rate of these files is $24$ kHz, resulting in signals with size $21,600,000$. To reduce this signal size, it is possible to arbitrarily break the $15$ minutes signal into smaller pieces, or to downsample the signal. The arbitrary separation of smaller pieces seem the least desirable approach, since it introduces the problem of deciding where to separate the pieces. 

For the following tests, however, no downsampling was adopted, and the reported results refer to the segmentation of the full $21,600,000$ points signal.

For the Bayesian binary algorithm with the Laplace prior, the selection of the $\beta$ value is done based on an elbow plot of the BIC criterion, i.e., we select the least $\beta$ for which the plot $BIC \times \beta$ shows a pronounced decrease. For the PELT algorithm, the MBIC criterion is applied. In the results in table \ref{tbl:real}, the execution time for the Bayesian binary algorithm with Laplace prior includes all the runs necessary to obtain the best $\beta$. In order to assess the effect of using strongly informative priors in our algorithm, we also included the results for the Bayesian binary algorithm using the Jeffreys' (non-informative) prior.

\begin{table*}\renewcommand{\arraystretch}{0.5}\caption{Results on real samples; see text for details}
\label{tbl:real}
\centering
\begin{tabular}{ccccccccc}
                  Sample &   Method &        Time (s) &      Beta &  True k &  Estimated k &  Precision &    Recall &        F1 \\
\midrule
 A &  jeffreys &    1239.59 &  - &      12 &         42074   &   0.03\% &  100\% &  0.0003 \\                                     
B &  jeffreys &    1329.73 &  - &      20 &           45277 &   0.04\% &  100\% &  0.0004 \\
\midrule
A &  laplace &    27.41 &  3.3e-5 &      12 &           28 &   17.9\% &  41.7\% &  0.1250 \\
B &  laplace &    30.89 &  1.6e-5 &      20 &           21 &   30.0\% &  30.0\% &  0.1500 \\
 \midrule
 A &     pelt &  205.41 &       - &      12 &        39170 &   0.03\% &  100\% &  0.0003 \\
 B &     pelt &  205.38 &       - &      20 &        38274 &   0.05\% &  100\% &  0.0005 \\
\end{tabular}
\end{table*}

As seen in table \ref{tbl:real}, the Bayesian binary algorithm showed superior results to PELT in the segmentation of real samples. The first thing to notice is that PELT resulted in an excessive number of changepoints; that is the case because PELT works with the exact optimization of a cost function that is based on a (Gaussian) likelihood, and even with the regularization induced with the MBIC criterion, a higher number of changepoints gives a better fit. The same happens with the Bayesian binary algorithm using non-informative priors, i.e., with uncontrolled power of the test based on the e-value. 

With the Bayesian binary algorithm, on the other hand, the value of $\beta$ helps to control the power of the test based on the e-value, avoiding oversegmentation.

\begin{figure*}
 \centering
 \includegraphics[width=0.9\textwidth]{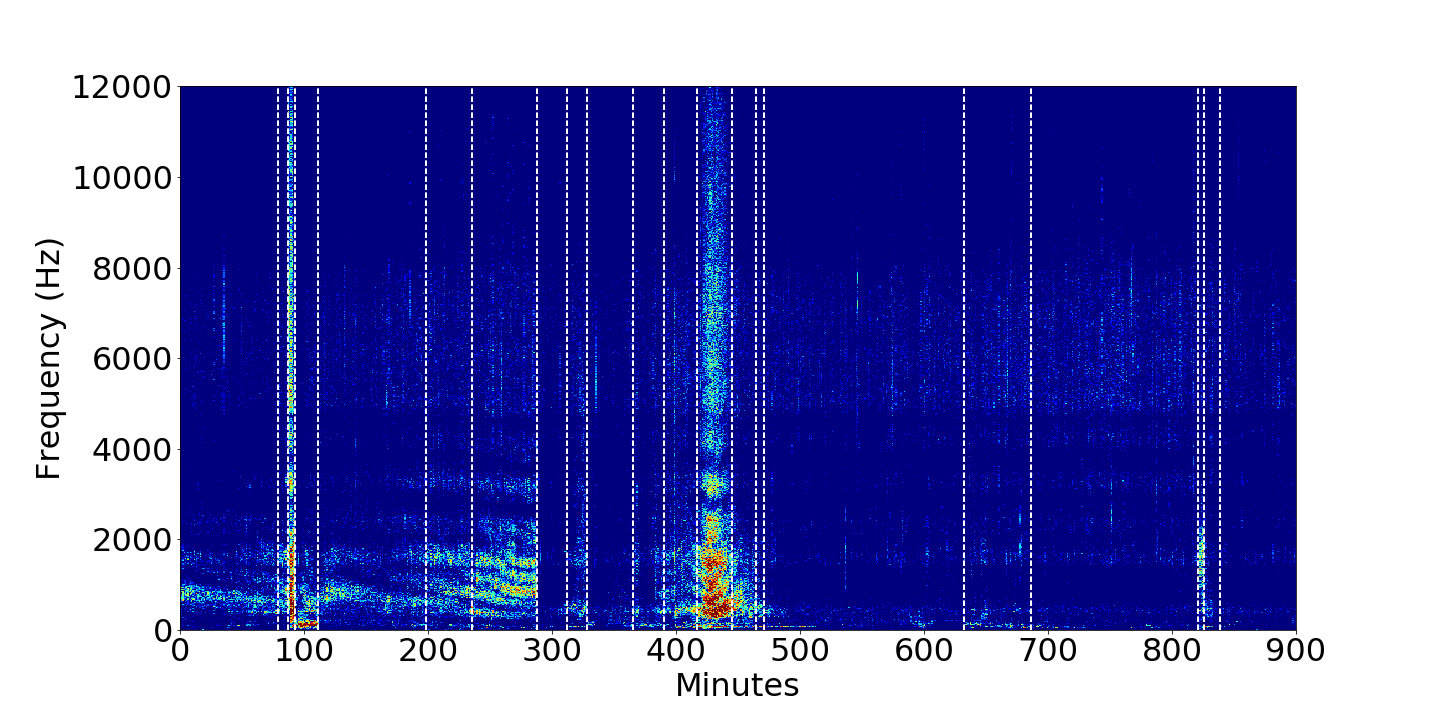}
 \caption{Spectrogram of sample A with changepoints estimated by the Bayesian binary algorithm}
 \label{fig:spec1}
\end{figure*}

\begin{figure*}
 \centering
 \includegraphics[width=0.9\textwidth]{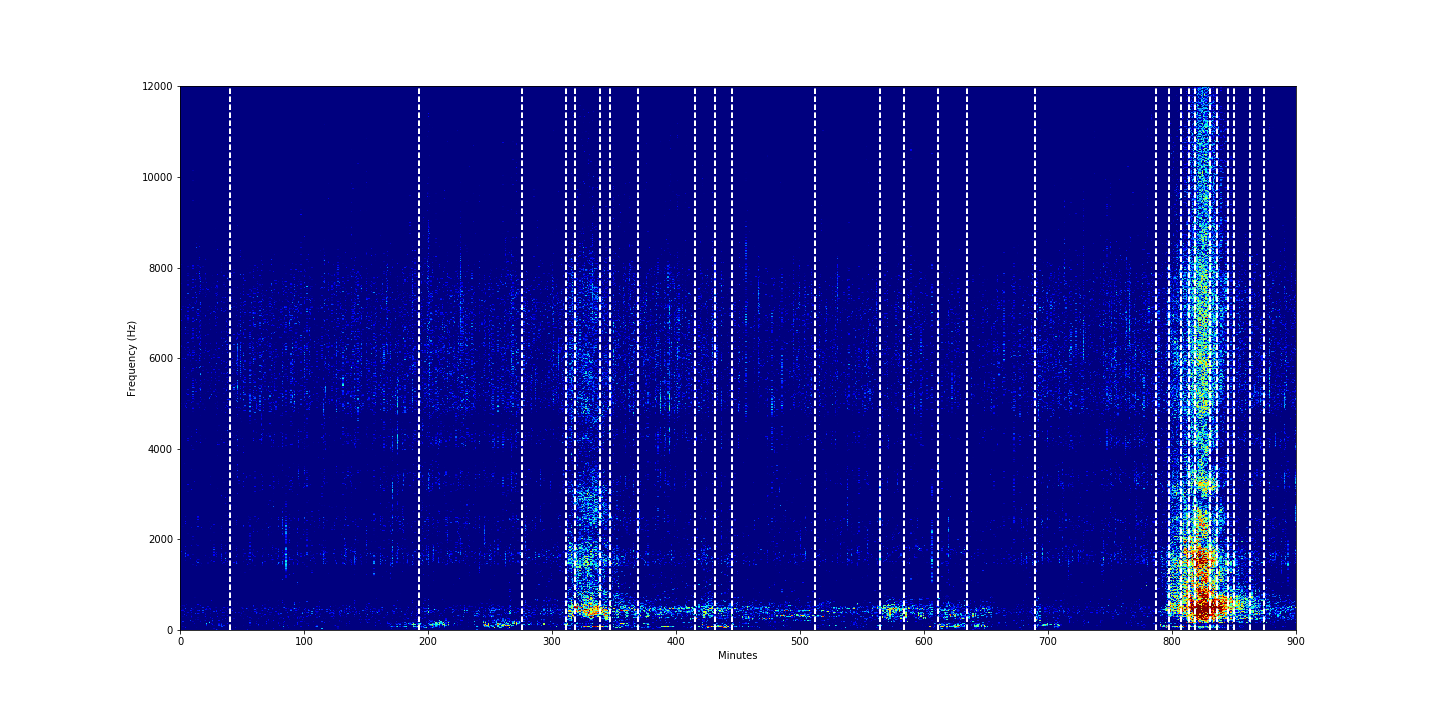}
 \caption{Spectrogram of sample B with changepoints estimated by the Bayesian binary algorithm}
 \label{fig:spec2}
\end{figure*}

In figures \ref{fig:spec1} and \ref{fig:spec2}, the changepoints estimated by the Bayesian binary algorithm are plotted over the spectrogram of the samples. It is noticeable that the boundaries of the most prominent events are correctly captured by the algorithm, while at the same time sections with no important events (as can be seen by direct inspection of the spectrogram) are kept unsegmented.

\section{Conclusion}\label{sec:conc}

The segmentation of acoustic signals is an important task, specially in the retrospective analysis of long duration signals. 

Among the many possible criteria for the segmentation, the RMS-based segmentation is particularly interesting when one is mainly interested in separating sections with background noise only, from sections composed of background noise plus some (possibly) interesting event.

In this paper, we present a Bayesian binary algorithm for RMS-based acoustic signal segmentation. We show that this algorithm is precise, and robust to violations on the basic assumptions: normality of background noise, and a stepfunction for the RMS in the different segments. We claim that this robustness is mainly due to two characteristics of our algorithm: first, the use of a marginal posterior for the selection of candidate changepoints; and second, the use of maximum entropy models (both the Gaussian for the background noise, and the Laplace for the log-ratio of variances are maximum entropy models) with strongly informative priors.

By comparing our algorithm with other alternatives from the literature, we showed that it is competitive with the current state-of-the-art changepoint algorithm (PELT), and sensibly superior to previous binary algorithms in simulated data. When analyzing real data, we showed that our algorithm can have superior results even when compared to PELT, if we use the strongly informative (Laplace) prior on the log-ratio of variances between segments. 

The hyperparameter of the Laplace prior can be efficiently selected using model selection criteria such as the Bayesian Information Criterion (BIC). 

Further work will analyze other possibilities for the model selection problem in this setting. We are also working on a hybrid version of our algorithm and the PELT algorithm, by using a version of our marginal posterior as the cost function to be optimized with PELT. 

Our algorithm is written in \emph{cython}, is open sourced an can be downloaded at \url{http://github.com/paulohubert/bayeseg}, along with some sample acoustic data and some illustrative \emph{IPython} notebooks. The signals used in this paper are available upon request.

\bibliographystyle{plain}
\bibliography{bibliografia}

\end{document}